\documentclass[prd, aps, superscriptaddress, twocolumn, preprintnumbers, floatfix, nofootinbib]{revtex4}
\usepackage[dvips]{graphicx}
\usepackage{epsf}
\usepackage{amsmath}
\usepackage{amssymb}

\usepackage{graphicx}
\usepackage{dcolumn}
\usepackage{bm}

\def\be{\begin{equation}}
\def\ee{\end{equation}}
\def\bea{\begin{eqnarray}}
\def\eea{\end{eqnarray}}

\usepackage{color}

\begin{document}


\title{Can Back-Reaction Prevent Eternal Inflation?}
\author{Robert Brandenberger$^{1}$, Renato Costa$^{1,2}$ and Guilherme Franzmann$^{}$ }
\email{rhb@physics.mcgill.ca, renatosa@ift.unesp.br, guilherme.franzmann@mail.mcgill.ca}

\affiliation{Department of Physics, McGill University, Montr\'eal, QC, H3A 2T8, Canada\\$^{2}$Instituto de F\'isica Teorica, UNESP-Universidade Estadual Paulista\\R. Dr. Bento T. Ferraz 271, Bl. II, Sao Paulo 01140-070, SP, Brazil}


\begin{abstract}

We study the effects which the back-reaction of long wavelength fluctuations exert on
stochastic inflation. In the cases of power-law and Starobinsky inflation these effects
are too weak to terminate the stochastic growth of the inflaton field. However, in the
case of the cyclic Ekpyrotic scenario, the back-reaction effects prevent the unlimited
growth of the scalar field.

\end{abstract}

\maketitle

\section{Introduction}

Inflationary cosmology \cite{Guth, Sato, Starob0, Brout} is the current
paradigm of early universe cosmology. According to this scenario, the
almost constant potential energy $V(\varphi)$ of a weakly coupled scalar field
$\varphi$ (the {\it inflaton}) leads to accelerated expansion of space. This accelerated
expansion of space redshifts all initial matter, leaving a vacuum state
behind. As first realized by Mukhanov and Chibisov for cosmological
fluctuations \cite{Mukh} and Starobinsky for gravitational waves \cite{Starob1},
quantum vacuum fluctuations are continuously generated on
sub-Hubble scales (wavelength smaller than the Hubble radius $H^{-1}(t)$,
where $H(t)$ is the cosmological expansion rate). As the wavelengths
of these fluctuation modes exit the Hubble radius, the vacuum oscillations
freeze out and the modes get squeezed (see e.g. \cite{MFB, RHBrev} for
reviews of the theory of cosmological perturbations) and become the
seeds for the observed inhomogeneities in the distribution of matter and
anisotropies in the cosmic microwave background (CMB) which are now being
mapped by cosmological experiments.

At quadratic order, the fluctuations effect the background space-time
and matter. The effects on the inflaton field lead to the
{\it stochastic inflation} scenario \cite{Starob2} which is one of the
cornerstones of the {\it inflationary multiverse} (see e.g. \cite{multi}
for a review) and has
possible implications for an anthropic ``resolution'' of the cosmological
constant problem (see e.g. \cite{anthro} for recent reviews
on dark energy and the cosmological constant problem).
Stochastic inflation is based on the effects
which the short wavelength cosmological fluctuations impart on the
effective background inflaton field when the wavelength of the fluctuation
modes cross the Hubble radius. The effects lead to a stochastic
source term in the equation of motion for $\varphi$ which leads to
an equal probability of driving $\varphi$ up or down the potential
$V(\varphi)$. Note that having $\varphi$ moving up its potential
does not violate covariant energy conservation since the covariant
conservation equations apply only to the sum of background and
fluctuations. Note also that
stochastic inflation is based on the infinite reservoir of ultraviolet
modes which are stretched by the accelerated expansion of space
taking place during the period of cosmological inflation.

At quadratic order, cosmological fluctuations and gravitational
waves also back-react on the metric. In inflationary cosmology,
these effects were first considered in the case of gravitational
waves in \cite{WT}, and in the case of cosmological perturbations
in \cite{ABM}. By the (almost) time translation symmetry of the
(almost) exponentially expanding background, the back-reaction
effects of sub-Hubble modes is constant in time. On the other
hand, since inflation builds up an increasing sea of super-Hubble
modes, the back-reaction effects of these infrared modes grows
in time. It was found \cite{ABM} that each long wavelength
modes leads to a small decrease in the effective energy
density. Note that it is the same infinite reservoir of fluctuation
modes which drives stochastic inflation which is responsible
for the increasing sea of long wavelength fluctuations modes
which can back-react.

Thus, in terms of the background energy density, there is a
competition between the effects of stochastic inflation which
lead (at least in half of space) to an increase in the energy
density, and those of back-reaction which lead to a decrease.
A natural question is whether back-reaction effects can
prevent stochastic inflation from continuing without bounds,
and hence prevent eternal inflation. This is the question
we here address in a couple of different scenarios:
{\it power-law inflation} \cite{Linde}, {\it Starobinsky inflation}
\cite{Starob0}, and also the phase of exponential expansion
taking place between the cycles of the {\it cyclic Ekpyrotic scenario}
\cite{ST}. We find that back-reaction effects are unable to
prevent eternal inflation for power-law and Starobinsky
inflation, but that they dominate over stochastic effects in
the cyclic Ekpyrotic scenario. The reason that back-reaction
effects dominate in the cyclic Ekpyrotic scenario is that
the fluctuations which are exiting the Hubble radius during
the accelerating phase are not in their vacuum state but have
already been highly squeezed during the phase of contraction.

In the following, we first summarize the basis of stochastic inflation,
followed by a review of back-reaction effects of super-Hubble
cosmological perturbations. We establish firstly the criterion
on values of $\varphi$ for which stochastic effects dominate over
the classical rolling of $\varphi$, and secondly the criterion for
back-reaction effects to dominate over the stochastic terms.
In Sections 4 - 6 we apply these conditions in turn to
power-law inflation, Starobinsky inflation, and to the phase
of accelerated expansion (the ``dark energy phase'') of the
cyclic Ekpyrotic scenario. We conclude with a summary of
our results and some discussions.

\section{Stochastic Dynamics of the Background Field}

Let us begin with a brief review of stochastic inflation \cite{Starob1}.
To put the discussion into context, consider the space-time
sketch of inflationary cosmology shown in Fig. 1. Here,
the horizontal axis gives the physical distance and the
vertical axis is physical time $t$. The inflationary phase
lasts from initial time $t_i$ until the final time $t_R$, the
time of {\it reheating}. The solid curve which is (almost)
vertical in the inflationary phase denotes the Hubble radius.
The dotted curves show the wavelengths of various
fluctuation modes which exit the Hubble radius during
inflation.

\begin{figure}
\includegraphics[scale=0.5]{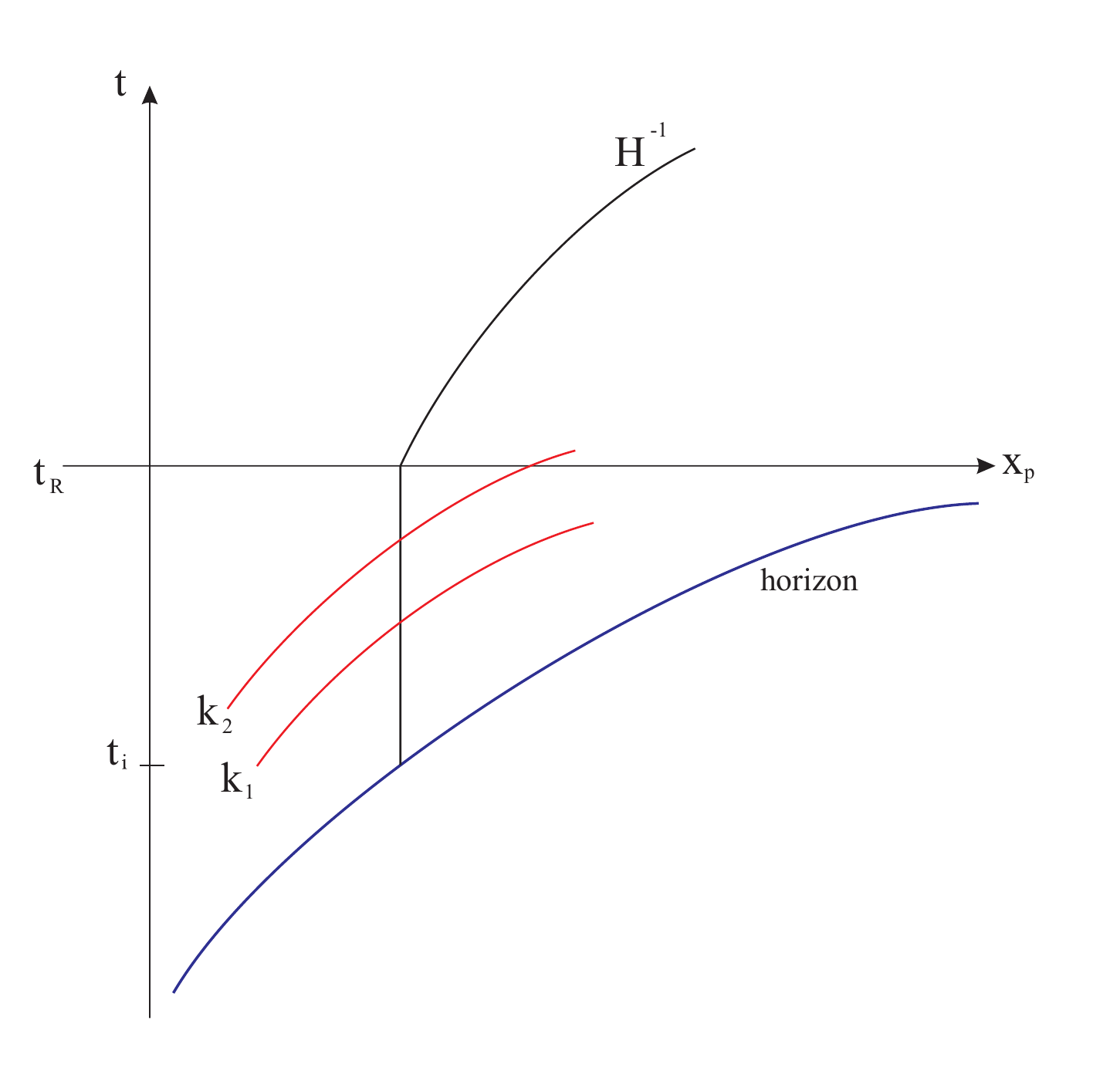}
\caption{Space-time sketch of an inflationary cosmology. The vertical axis is time, the horizontal axis is physical distance. The inflationary phase lasts from $t_i$ until $t_R$, and it is during this period that the fluctuations we consider here exit the Hubble radius.}
\end{figure}

The formalism of stochastic inflationary dynamics describes
the effect of modes exiting the Hubble radius on the
evolution of the effective background field which is the full
field coarse grained over the Hubble volume (see e.g.
\cite{Laurence} for a modern view on the formalism of
stochastic inflation). In slight abuse of notation we
will also denote the effective background field (and not
just the full field) by $\varphi$. Taking into account
the effects of modes crossing the Hubble radius which
are now entering the sea of long wavelength modes,
the equation of motion for $\varphi$ becomes
\be \label{EoM}
{\ddot{\varphi}} + 3 H {\dot{\varphi}} + V^{\prime}(\varphi) \, = \, \frac{3}{2 \pi} H^3 \xi(t)  \, ,
\ee
which assumes that $H$ is approximately constant and where the prime
denotes the derivative with respect to
$\varphi$, and where $\xi(t)$ is a Gaussian random variable with unit variance
which takes on different values in different Hubble patches.
If $\xi(t)$ is positive, then the source term in (\ref{EoM}) will
drive $\varphi$ up the potential, but if $\xi(t)$ is negative, then
the stochastic source will reinforce the classical force driving
$\varphi$ down its potential.

The {\it stochastic region} of field space is defined to be
the one for which the classical force term (the right hand side
of (\ref{EoM}) exceeds the classical force in magnitude, i.e.
\be  \label{StochCond}
\frac{3}{2 \pi} H^3 \, \geq \, |V^{\prime}(\varphi)| \, .
\ee
For example, for the simplest chaotic inflation model with
potential
\be
V(\varphi) \, = \, \frac{1}{2} m^2 \varphi^2
\ee
the condition (\ref{StochCond}) becomes
\be \label{StochCond1}
|\frac{\varphi}{m_{pl}}| \, \geq \, \sqrt{4 \pi} 6^{1/4} \biggl( \frac{m_{pl}}{m} \biggr)^{1/2} \, ,
\ee
where $m_{pl}$ is the reduced Planck mass defined in terms of
Newton's constant $G$ via $m_{pl} = ( \sqrt{8 \pi G} )^{-1}$.

Note that for the normalization of the mass $m$ which is consistent
with the observed amplitude of CMB anisotropies, the stochastic
region of field space is far beyond the field values which influence
the period of inflation which is observationally accessible to us. They
do, however, correspond to energy densities which are still
much lower than Planck densities.

\section{Back-Reaction of Long Wavelength Fluctuations}

In early universe cosmology we usually consider a homogenous
and isotropic background space-time and superimpose small
amplitude cosmological fluctuations which are treated by linearizing
the field equations about the background. The Einstein
field equations, however, are highly nonlinear, and hence
even at the classical level the fluctuations at second order
influence the background. This is what we mean by
{\it back-reaction}.

The expansion parameter for cosmological perturbation theory
is the amplitude of the fluctuations which is set by the
size of the observed CMB anisotropies and is of the
order $10^{-5}$. Hence, the back-reaction effect of any
given fluctuation mode is tiny (of the order $10^{-10}$).
However, each fluctuation mode effects the background,
and hence, for a long period of inflation the integrated
effect of all of the modes can be important (see \cite{RHBrev1}
for a review of back-reaction effects of long wavelength
cosmological perturbations).

As mentioned in the introductory section, the back-reaction
effect of sub-Hubble fluctuations is very small and approximately
constant in time. On the other hand, the back-reaction of
long wavelength modes shows secular growth due to the
increasing phase space of super-Hubble modes.

How is it possible that super-Hubble scale modes can effect
local physics? First of all, note that there is no causality
obstacle. During inflation, the causal horizon (the forward
light cone) is (almost) exponentially stretched compared
to the Hubble radius. All modes we are considering have a
wavelength which is smaller than the horizon. Secondly, note
an analogy with black hole physics \footnote{One of us (R.B.)
is grateful to Richard Woodard for making this point.}.
If we throw a mass into a black hole, then the mass may
have forever disappeared beyond the horizon, but the
gravitational effects of this mass remain visible to the
observer at infinity. In a similar way, the local gradients
of a long wavelength cosmological fluctuation mode may
decay exponentially, but the gravitational effects of this
mode will persist and have a local effect.

The gravitational back-reaction formalism is as follows \cite{ABM}.
We begin with the full Einstein field equations
\be \label{Einstein}
G_{\mu \nu} \, = \, 8 \pi G T_{\mu \nu} \, ,
\ee
where $G_{\mu \nu}$ is the Einstein tensor and $T_{\mu \nu}$ is
the energy-momentum tensor of matter. The background
metric $g^{(0)}_{\mu \nu}(t)$ and matter $\varphi^{(0)}(t)$
obey these (nonlinear) equations. Next we introduce the
linearized metric and matter fluctuations $\delta g_{\mu \nu}$
and $\delta \varphi$ which are both functions of space and
time and which obey the linearized Einstein equations. However,
the system of fields
\bea \label{ansatz}
g_{\mu \nu}(x, t) \, &=& \, g^{(0)}_{\mu \nu}(t) + \delta g_{\mu \nu}(x, t) \, \nonumber \\ \varphi(x, t) \, &=& \, \varphi^{(0)}(t) + \delta \varphi(x, t) \,
\eea
does not satisfy the Einstein equations at second order.

If we are only interested in modifying the background at
quadratic order (modifications of the fluctuations can be
considered as well \cite{Patrick}) we need to introduce
second order correction terms $g^{(2)}_{\mu \nu}(t)$
and $\varphi^{(2)}(t)$ to metric and matter.
Adding these terms to the ansatz (\ref{ansatz}) for
metric and matter, inserting into the Einstein equations (\ref{Einstein}),
cancelling the linear terms using the linear fluctuation
equations, moving all terms quadratic in the
fluctuations to the right hand side, and taking the spatial
average of the resulting equation yields an
equation of motion for the corrected background metric
\be
g^{(0, br)}_{\mu \nu}(t) \, = \, g^{(0)}_{\mu \nu}(t) + g^{(2)}_{\mu \nu}(t)
\ee
of the following form:
\be
G_{\mu \nu}(g^{(0. br)})(t) \, = \, 8 \pi G \biggl[ T^{(0)}_{\mu \nu}(\varphi^{(0)}(t)) + \tau_{\mu \nu}(t) \biggr] \, ,
\ee
where $\tau_{\mu \nu}(t)$ is quadratic in the cosmological fluctuations, and is
called the ``effective energy-momentum tensor of cosmological perturbations''.
The effective energy momentum tensor is obtained by integrating over the
contributions of all fluctuation (Fourier) modes. For the reasons explained above
we are only interested in the contribution of the super-Hubble modes.

As background metric we take a spatially flat Friedmann-Robertson-Walker-Lem\^aitre
metric given by the line element
\be
ds^2 \, = \, a(\eta)^2 \bigl( d\eta^2 - d\vec{x}^2 \bigr) \, ,
\ee
where $\eta$ is conformal time. To evaluate $\tau_{\mu \nu}$ we work in
longitudinal (conformal-Newtonian) gauge in which (for single component matter
without anisotropic stress) the line element is
\be
ds^2 \, = \, a(\eta)^2 \biggl[ (1 + 2 \Phi) d\eta^2 - (1 - 2 \Phi) d\vec{x}^2 \biggr] \, ,
\ee
where $\Phi(x, t)$ is the relativistic gravitational potential \cite{Bardeen}.

The metric and matter fluctuations are not independent. They are coupled
via the Einstein constraint equations. In the background of a slowly rolling
scalar field the connection is given by
\be
\delta \varphi(k) \, = \, -\frac{2 V}{V^{\prime}} \Phi(k) \, ,
\ee
where the argument $k$ indicates that we are considering the
Fourier modes of the fluctuations. The general expression for
$\tau_{\mu \nu}$ contains many terms (see e.g. \cite{RHBrev1}
for the full expression). However, for super-Hubble modes the
terms containing spatial and temporal derivatives can be neglected,
the latter since in an expanding universe the dominant mode of
$\Phi$ is constant in time. The terms which survive give a
contribution to the energy density of the form
\be \label{rhobr}
\rho_{br}(t) \, \simeq \, \biggl[2\frac{V^{\prime\prime}V^2}{(V^{\prime})^2}-4V\biggr] \langle \Phi^2 \rangle \, ,
\ee
where $\langle \Phi^2 \rangle$ is obtained by integrating over all of the super-Hubble modes.
In the simple chaotic inflation model considered in \cite{ABM}, the second
term in (\ref{rhobr}) is larger in magnitude than the first, and hence the
effective energy density of long wavelength cosmological perturbations is
negative. We shall see that the same is true in the other models considered
here.  The effective pressure of cosmological perturbations is
\be
p_{br} \, \simeq \, - \rho_{br} \, ,
\ee
and hence long wavelength cosmological fluctuations effect the background
geometry like a negative cosmological constant (with possible implications for
a possible solution of the cosmological constant problem discussed in
\cite{RHBrev1}). The physical reason why long wavelength fluctuations
act as a negative cosmological constant is easy to understand: a matter
fluctuation (with positive matter energy density) leads to a potential well
(negative gravitational energy density), and on super-Hubble scales the
magnitude of the gravitational energy is larger than that of the matter energy,
hence leading to a negative effective energy density. Since no terms with
spatial and temporal gradients contribute, the equation of state of $\tau_{\mu \nu}$
has to be that of a cosmological constant.

An important question to ask \cite{Unruh} is whether the effects of the
contribution of super-Hubble modes to $\tau_{\mu \nu}$ are locally measurable.
Returning to the black hole example, we note that there needs to be an
observer (the observer at infinity) to measure the change in the mass of the
black hole when some matter is thrown into it. In a similar way, a physical
clock field is required in order to locally measure the effects of the long
wavelength contribution to $\tau_{\mu \nu}$. For purely adiabatic fluctuations,
the effect is equivalent \cite{Ghazal1, AW} to a second order time-translation.
However, in terms of a clock field $\chi$, the effects of the back-reaction
of super-Hubble modes is physically measurable \cite{Ghazal2}, and it
corresponds to a decrease in the local Hubble expansion rate \cite{Marozzi}.
From now on we will implicitly assume that we have a clock field present
which plays the same role as the CMB plays in late time cosmology in
setting the clock without producing curvature of space.

At this point we have established that stochastic effects lead (at least in half
of space) to an increase in the energy density, whereas the back-reaction
of super-Hubble modes leads to a decrease. In the following, we will
compare the magnitude of these effects in various cosmological models.

\section{Case 1: Power-Law Inflation}

We first consider large field power-law inflation models with
potential
\be \label{power}
V(\varphi) \, = \, \lambda m^{4 - n} \varphi^n \, ,
\ee
where $\lambda$ is a dimensionless constant and $m$ is a
mass scale. This class of potentials include the
simple chaotic inflation models with $n = 2$ and $n = 4$,
and axion monodromy inflation models which $n$ can
be a real number in the range $0 < n < 2$ \cite{axion}.
In the case of $n \neq 4$ we can without loss of
generality set $\lambda = 1$. However, sometimes
it is convenient to keep $\lambda$ and instead
replace $m$ by $m_{pl}$. With the first choice, the
condition to obtain small density fluctuations is $m \ll m_{pl}$,
in the second case $\lambda \ll 1$. Note that the region
of slow roll inflation corresponds to trans-Planckian field
value
\be
|\varphi| \, > \, \alpha(n) m_{pl} \, ,
\ee
where the $\alpha(n)$ is of the order $1$.

Taking the derivative of (\ref{power}) to obtain the
classical force and comparing with the stochastic force
given by the right hand side of (\ref{EoM}) we get
the following field range for which the stochastic
force dominates over the classical one:
\be \label{StochCond1}
| \varphi | \, > \, \biggl( 2 \pi \sqrt{3} n \biggr)^f \lambda^{-f/2} \biggl( \frac{m_{pl}}{m} \biggr)^{(2 - \frac{1}{2}n)f} m_{pl} \, ,
\ee
where the exponent $f$ is
\be
f \, = \, \frac{1}{n/2 + 1} \, .
\ee
It is easy to see that in the case $n = 2$ and $\lambda = 1/2$ we
recover the condition (\ref{StochCond}).

If we are in a region of space in which stochastic effects
drive $\varphi$ up the potential, the increase in the potential
energy over one Hubble time $H^{-1}$ is then given by
\be \label{deltaV}
\Delta V \, \simeq \, \Delta \varphi V^{\prime} \, ,
\ee
where
\be \label{deltavarphi}
\Delta \varphi \, = \, \frac{H}{2 \pi}
\ee
is the change in $\varphi$ over one Hubble time step (the
coefficient in (\ref{deltavarphi}) is consistent with the
coefficient of the stochastic term in (\ref{EoM})).

As discussed in the previous section, the back-reaction
of super-Hubble modes leads to a negative contribution to
the energy density which grows in time during a period
of accelerated expansion since modes are exiting the Hubble
radius and increasing the sea of infrared modes. In order
to compare the change in the energy density due to back-reaction
with the change due to stochastic evolution we need to
evaluate the change in $\langle \Phi^2 \rangle$ over a Hubble time,
the same time interval considered above for the stochastic
effect.

The starting point is the following expression for
the contribution to $\langle \Phi^2 \rangle$ from super-Hubble
Fourier modes $\Phi(k)$:
\be
\langle \Phi^2 \rangle(t) \, = \, 4 \pi \int_0^{k_H(t)} k^2 |\Phi(k)|^2 dk\, ,
\ee
where $k_H(t)$ is the comoving wavenumber corresponding
to Hubble radius crossing at time $t$. The change
in $\langle \Phi^2 \rangle$ over one Hubble time is then given
by
\be
\Delta \langle \Phi^2 \rangle \, = \, H^{-1} 4 \pi k_H^2 |\Phi(k_H(t))|^2 \frac{dk_H(t)}{dt} \, .
\ee
In the case of an exponentially expanding background we
have
\be
k_H(t) \, = \, a(t) H \, .
\ee
The corrections to this formula in the case of slow-roll
inflation are negligible.

In the case of inflation, the value of $\Phi(k)$ at
Hubble radius crossing is given by the vacuum initial
conditions \cite{Mukh, RHBrev}. We use the relations
\bea
\zeta(k) \, &\simeq& \, \frac{2}{3} \frac{1}{1 + w} \Phi(k) \nonumber \\
\zeta(k) \, &=& \, z^{-1} v(k) \, ,
\eea
which express the gravitational potential $\Phi$ in terms
of the curvature fluctuation variable $\zeta$ (which is
conserved in an expanding universe on super-Hubble scales),
which is then in turn related to the canonical fluctuation
variable $v$ \cite{Sasaki, Mukh2} via the background variable $z$
which is given by
\be
z \, = \, \frac{a {\dot{\varphi}^{(0)}}}{H} \, .
\ee
In the above, the equation of state parameter $w$ is the
ratio of pressure to energy density. Using vacuum initial
conditions for $v(k)$ we then obtain
\be \label{deltaphi}
\Delta \langle \Phi^2 \rangle \, = \, \frac{9}{2} \pi H^4 \frac{(\dot{\varphi}^{(0)})^2}{V^2} \, .
\ee
Inserting back into the expression (\ref{rhobr}) for the effective
energy density of back-reaction we then obtain (after using
the slow-roll equation of motion to replace $\dot{\varphi}^{(0)}$
in terms of $V^{\prime}$ and $V$):
\be \label{rhobr2}
\Delta \rho_{br} \, = \, \frac{\pi}{3} \biggl[ V^{\prime \prime} V - 2 (V^{\prime})^2 \biggr] m_{pl}^{-2} \, .
\ee

We are now able to compare the magnitude of the increase in
energy density due to stochastic rolling up the potential with
the decrease due to the increase in $\rho_{br}$. Combining
the above equations (\ref{deltaV}), (\ref{deltavarphi}) and (\ref{rhobr2})
yields
\be \label{ratio}
\frac{|\Delta \rho_{br}|}{\Delta V} \, = \, \frac{2 \pi^2}{\sqrt{3}} \sqrt{\lambda} m^{2 - \frac{1}{2}n}(n + 1) \varphi^{\frac{1}{2}n - 1} m_{pl}^{-1} \, .
\ee
Without loss of generality we can set $m = m_{pl}$ and represent
the small slope of the potential (which is required in order
for the cosmological fluctuations induced by inflation not to
exceed the observational upper bound) by a small value
$\lambda \ll 1$ of the coupling constant.

In the case $n = 2$ it is clear from (\ref{ratio}) that $|\Delta \rho_{br}|$
is smaller than $\Delta V$ for all field values. Hence, back-reaction
cannot prevent eternal inflation. For $n > 2$ there is a critical
field value $\varphi_c$ beyond which $|\delta \rho_{br}|$ exceeds
$\Delta V$:
\be
\varphi_c \, = \, \biggl( \frac{\sqrt{3}}{2 \pi^2} \frac{1}{n + 1} \biggr)^{\frac{1}{n/2 - 1}} \lambda^{-\frac{1}{n - 2}} m_{pl}
\ee
which corresponds to trans-Planckian energy densities. Once again,
back-reaction cannot prevent eternal inflation. Finally, for $n < 2$ the
exponents reverse sign and the condition for $\Delta \rho_{br}$
to dominate becomes an upper bound for $|\varphi|$:
\be
|\varphi|^{\tilde{n}} \, < \, \frac{1}{3} \lambda^{1/2} (n + 1) m_{pl}^{\tilde{n}} \, ,
\ee
where ${\tilde{n}} \equiv 1 - n/2$. This is not the field range for inflation.

In conclusion, we find that in no version of simple power law inflation
models back-reaction of long wavelength fluctuations can prevent
eternal inflation.

\section{Case 2: Starobinsky Inflation}

Starobinsky's initial model of exponential expanding space was
based on a higher derivative gravitational Lagrangian \cite{Starob0}.
After a conformal transformation, it corresponds to Einstein gravity
in the presence of a scalar matter field $\varphi$ with exponential
potential
\be
V(\varphi) \, = \, A \bigl( 1 - e^{-b \varphi} \bigr)^2 \, ,
\ee
where the $A \ll m_{pl}^4$ and $b \sim m_{pl}^{-1}$. Such potentials
also arise in chaotic inflation in supergravity \cite{GL}.

The region of inflation once again corresponds to trans-Planckian
field values where the potential energy is approximately
given by $A$. As in the case of power law inflation discussed
in the previous section, we first determine the field range
where stochastic effects dominate. Demanding that the
stochastic force amplitude exceeds the classical force
yields the condition
\be
\varphi \, > \, b^{-1} {\rm ln} \biggl[ \frac{3}{4 \pi} \biggl( \frac{1}{3} \biggr)^{3/2} \biggl( \frac{A}{m_{pl}^4} \biggr)^{1/2} (b m_{pl})^{-1} \biggr] \, .
\ee

Now we can turn to a comparison of the increase in potential
energy due to stochastic rolling up the potential to the
change in the energy density of back-reaction.
Making use of (\ref{deltaV}) and (\ref{deltavarphi}),
expressing $H$ in terms of $V$, and making the approximation
$V \simeq A$ we obtain
\be
\Delta V \, \simeq \, \frac{1}{\sqrt{3} \pi} b A^{3/2} e^{-b \varphi} m_{pl}^{-1} \, .
\ee
On the other hand, from (\ref{rhobr}) and (\ref{deltaphi}), the change in the energy density
of back-reaction is given by
\be
\Delta \rho_{br} \simeq \, - \frac{2 \pi}{3} A^2 b^2 e^{-b \varphi} m_{pl}^{-2} \, .
\ee

The ratio is
\be
\frac{|\Delta \rho_{br}|}{\Delta V} \, = \, \frac{2 \sqrt{3}}{3} \pi^2 A^{1/2} b m_{pl}^{-1}
\ee
which is much smaller than unity for the values of $A$ and $b$ which
need to be chosen to get successful inflation. Hence, we conclude
that also in Starobinksy inflation back-reaction terms are too weak
to prevent eternal inflation.

\section{Case 3: Cyclic Ekpyrotic Scenario}

Finally, we turn to the ``dark energy phase'' of the cyclic
Ekpyrotic scenario. The Ekpyrotic scenario is an alternative
to inflation for producing the observed inhomogeneities and
anisotropies \cite{Ekp}. It is based on the Horava-Witten scenario
of heterotic M-theory \cite{Horava}, a higher dimensional model.
At the effective field theory level it reduces to the theory
of a scalar field (in the higher dimensional picture
it corresponds to the separation of parallel branes)
coupled to Einstein gravity. The potential of the scalar field
is argued to be a negative exponential. This setup leads
to a bouncing cosmology.
According to the Ekpyrotic scenario, the universe begins in a phase
of contraction in which the  scalar field  is rolling down the
potential. Since the potential is negative, one obtains
an equation of state with $w \gg 1$, where the equation of state
parameter $w$ is the ratio of energy density and pressure.
Once the scalar field drops below $\varphi = 0$ (which corresponds
to the brane separation approaching the string scale), a
cosmological bounce is assumed to take place during which
regular matter and radiation are produced, leading to a Standard
Big Bang phase of expansion during which $\varphi$ climbs
back up the potential (while being a subdominant form of matter).
Cosmological fluctuations are created during the phase of
contraction. As long as an almost massless entropy field is
present (and this completely natural from the higher dimensional
point of view \cite{BBP}), an almost scale-invariant spectrum of
curvature perturbations is generated
\cite{Notari, Turok, Creminelli, NewEkp}.

By introducing a slight lift of the potential, i.e. by choosing
\be \label{ExpPot}
V(\varphi) \, = \, C - V_0 e^{-a \varphi} \, ,
\ee
the Ekpyrotic scenario becomes ``cyclic'' \cite{ST}. Once $\varphi$
during the phase of cosmic expansion reaches values with
$V(\varphi) > 0$, a period of accelerated expansion will
start. The scenario thus includes dark
energy. Based on the classical equation of motion for
$\varphi$ we would conclude that $\varphi$ will eventually
turn around and start to decrease. This leads to a
phase of contraction: the Ekpyrotic scenario has become
cyclic. Figure 2 presents a sketch of space-time
in the Ekpyrotic scenario. The vertical axis is time,
the horizontal is comoving distance. The bounce
time is taken to be $t = 0$. The Hubble radius
decreases very fast before the bounce and then
slowly increases in the Standard Big-Bang phase
after the bounce. For the cyclic Ekpyrotic scenario to work the
way it is meant to, the constant $C$ corresponds to
the currently observed cosmological constant, whereas
the value of $V_0$ corresponds typically to a very
high scale (e.g. the energy scale of Grand Unification).
Also, the value of $a$ will be given by the inverse string scale.

\begin{figure}
\includegraphics[scale=0.5]{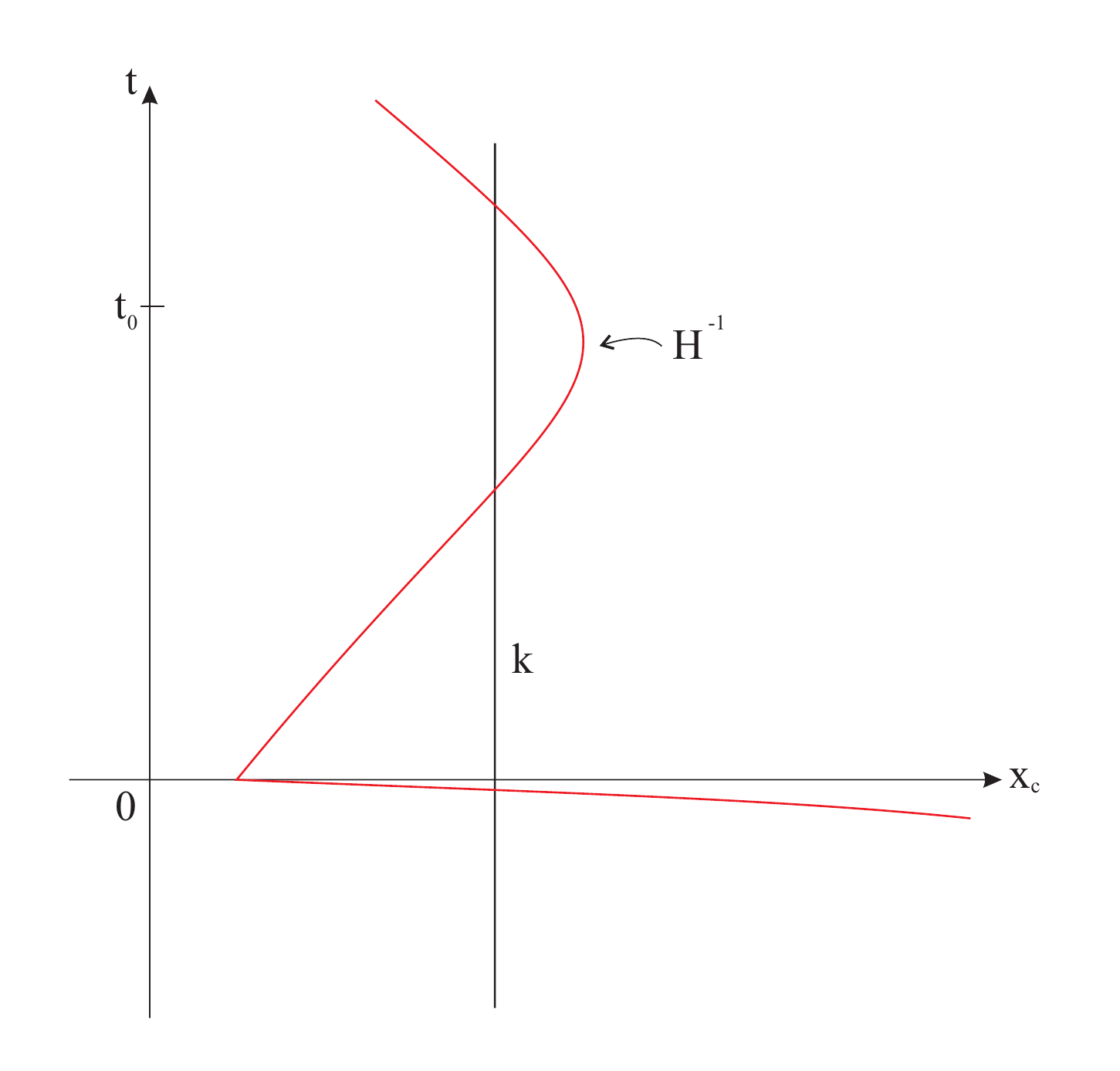}
\caption{Space-time sketch of the cyclic Ekpyrotic scenario. The vertical axis is time,
the horizontal is comoving physical distance. Fluctuations are generated during the
phase of contraction for $t < 0$. The bounce occurs at $t = 0$ and is followed by an
expanding phase. Close to the present time, we enter the dark energy phase driven
by the term $C$ in the potential (\ref{ExpPot}). The fluctuations which exit the Hubble
radius during this phase are not vacuum perturbations, but the evolved ones and
have a much larger amplitude than vacuum fluctuations would have.}
\end{figure}

A question to ask is whether stochastic effects analogous
to the ones which drive stochastic eternal inflation will
lead to an eternal stochastic growth of $\varphi$ in
the Ekpyrotic scenario, thus leading to an ``Ekpyrotic
landscape''. Intuitively we would not expect this to happen
since the stochastic effects are highly suppressed in
the dark energy phase (since $C <<< m_{pl}^4$) whereas
the cosmological fluctuations (and hence their back-reaction
effects) are not suppressed compared
to the case of inflation. In fact, when the fluctuations
exit the Hubble radius in the accelerating phase, they are
not in the vacuum state, unlike in the case of Starobinsky
inflation. Hence, the formula for the energy density of
back-reaction is different. These effects lead to an
enhancement of the back-reaction ``force'' compared to
the stochastic one. In the following we will show that
these expectations are indeed borne out.

The value of $\varphi$ which corresponds to $V = 0$ will be
denoted by $\varphi_0$ and is given by
\be
e^{- a \varphi_0} \, = \, \frac{C}{V_0} \, .
\ee
For field values significantly larger than $\varphi_0$ we can
approximate the value of the potential by $V = C$. In this
case, the increase in potential energy while rolling up the
potential for one Hubble time is given by
\be \label{EkpDeltaV}
\Delta V \, \simeq \, \frac{1}{2 \pi} \biggl( \frac{1}{3} \biggr)^{1/2} C^{3/2} a m_{pl}^{-1} e^{-a \delta \varphi} \, ,
\ee
where
\be
\varphi \, \equiv \, \varphi_0 + \delta \varphi \, .
\ee

Turning to the evaluation of the change in the energy density
due to back-reaction, it is important to note that the fluctuations
which are exiting the Hubble radius during the dark energy
phase are not the vacuum ones, but the ones which have evolved
and have produced the structure which we see on large scales. We
hence have
\be
\Delta \langle \Phi^2 \rangle \, \sim \, 1
\ee
and hence from (\ref{rhobr})
\be
\Delta \rho_{br} \, = \, 2 \biggl[ \frac{V^{\prime \prime} V^2}{(V^{\prime})^2} - 2 V \biggr] \, .
\ee
Inserting the form of the Ekpyrotic potential and considering large
field values we find
\be \label{EkpBR}
\Delta \rho_{br} \, \sim \, - 2 C e^{a \delta \varphi} \, .
\ee
Comparing (\ref{EkpDeltaV}) and (\ref{EkpBR}) we find
\be
\frac{|\Delta \rho_{br}|}{\Delta V} \, = \, 4 \pi \sqrt{3} \frac{a^{-1} m_{pl}}{C^{1/2}} e^{2 a \delta \varphi} \, .
\ee
Since the scale $C$ corresponds to the current dark energy scale, the
coefficient in front of the exponential in the above equation is
many orders of magnitude larger than $1$. Hence we conclude that
in the cyclic Ekpyrotic scenario back-reaction prevents the
stochastic growth of $\varphi$ and that there hence is no
eternal expansion in the late time dark energy phase for this model.
\\
\section{Conclusions and Discussion}

Stochastic effects will lead to the effective scalar field
climbing up the potential in some regions of space. This leads to
an increase in the energy density. On the other hand,
the back-reaction of fluctuations which have already exited
the Hubble radius will lead to a decrease in the effective
energy density. In this paper we have compared the magnitude
of the two effects in various cosmologies with an accelerating
phase.

We have shown that for both power-law and Starobinsky inflation
that strength of the back-reaction effect is too weak to prevent
the stochastic growth of $\varphi$ and hence does not cut off
eternal inflation. On the other hand, the back-reaction of
fluctuations in the dark energy phase of the cyclic Ekpyrotic
scenario greatly overwhelms the increase in energy due to
stochastic dynamics, and hence no eternal expansion in the late
time dark energy phase for the Ekpyrotic scenario is generated.

It is worth mentioning that besides the calculations done in this paper,
we have also looked into generalizations of them. The first idea was to
consider that $N$ Hubble times could pass so that more modes could leave
the Hubble radius and the back-reaction effect would be increased to the
point that even power-law and Starobinsky models could not have
eternal inflation. On the other hand, this would also enhance the stochastic
effect since the total $\Delta\varphi$ would be larger. At the end, we
found that the number $N$ for which back-reaction starts to overwhelm
the stochastic effect would be unreasonably large in the setup of inflation.

Furthermore, one could notice that (\ref{rhobr}) assumes the slow-roll
equation of motion for the inflaton field. However, since the field is rolling
under the influence of the stochastic force,
this formula should be generalized. Therefore, we have also treated all
the scenarios discussed in this paper
using the generalization of the effective energy density of back-reaction
obtained when considering the correct equations of motion.
We found that 1) for power-law inflation, there are field regions in which
the back-reaction effective energy density
becomes positive instead of negative, thus being completely unable to
prevent eternal inflation;
2) for Starobinsky inflation and for the Ekpyrotic scenario, the conclusions
remain the same as the ones obtained here.

It is also worth while commenting on the relation of our work to the
conclusions reached in \cite{ABM}. In those works, the question addressed
was what the absolute magnitude of the back-reaction energy density
is assuming that slow-roll inflation starts at some field value $\varphi_i$
and lasts until $\varphi \sim m_{pl}$. In the case of a potential
$V(\varphi) = \frac{1}{2} m^2 \varphi^2$ it was found that if
$\varphi_i > m^{-1/3}$ (in Planck units) then the total energy in the
back-reacting infrared modes becomes larger than the background
energy density. In that work, however, the effects of the stochastic
noise leading to an increase in $\varphi$ was not considered.
Thus, the questions considered in \cite{ABM} and in this paper are
complementary.
\\
\acknowledgements{We wish to thank I. Morrison for valuable discussions,
and P. Steinhardt for comments on the manuscript.
RB is supported by an NSERC Discovery Grant, and by funds from the Canada
Research Chair program. RC is supported by CAPES (Proc. 5149/2014-02)
and GF is supported by CNPq through the Science without Borders (SwB).}

\end{document}